\documentclass[]{spie}  %>>> use for US letter paper
\usepackage{amsmath,amsfonts,amssymb}
\usepackage{graphicx}
\usepackage[colorlinks=true, allcolors=blue]{hyperref}

\title{Measuring intensity correlations of a THz quantum cascade laser around its threshold at sub-cycle timescales}

\author[a]{Ileana Cristina Benea Chelmus}
\author[a]{Christopher Bonzon}
\author[a]{Curdin Maissen}
\author[a]{Giacomo Scalari}
\author[a]{Mattias Beck}
\author[a]{J\'er\^ome Faist}
\affil[a]{ETH Zurich, Institute of Quantum Electronics, Auguste-Piccard-Hof 1, Zurich 8093, Switzerland}

\authorinfo{Further author information: (Send correspondence to I.C.B.C)\\I.C.B.C: E-mail: ileanab@ethz.ch, Telephone: +41 44 633 76 36}

\begin{document} 
\maketitle
\begin{abstract}
The quantum nature of photonic systems is reflected in the photon statistics of the light they emit. Therefore, the development of quantum optics tools with single photon sensitivity and excellent temporal resolution is paramount to the development of exotic sources, and is particularly challenging in the THz range where photon energies approach k$_{b}$T at T=300 K. Here, we report on the first room temperature measurement of field $g^{1}(\tau)$ and intensity correlations $g^{2}(\tau)$ in the THz range with sub-cycle temporal resolution (146 fs) over the bandwidth 0.3-3 THz, based on electro-optic sampling. With this system, we are able to measure the photon statistics at threshold of a THz Quantum Cascade Laser.   
\end{abstract}


\section{INTRODUCTION}
\label{sec:intro}  % \label{} allows reference to this section

The Terahertz frequency range of the electromagnetic spectrum~(0.1-10~THz), lying between the optics and electronics, has witnessed up to date very advanced applications in spectroscopy and matter control~\cite{Kampfrath2013}, despite technological challenges which often require cryogenics for low background noise and free-space optics for minimized absorption. Devices operating in this range generally combine technologies belonging both to optics and to high-frequency electronics, which lead to light emission by uncommon mechanisms: lasers based on intersubband transitions configured with metallic strip wave-guides, photo-conductive antennas based on charge transients just to name a few. Nevertheless, the complexity of optical systems and the variety of applications is thus hindered by technological challenges, with notable realizations in the field of afm-imaging~\cite{EiseleM.2014}, higher harmonics generation~\cite{Hohenleutner2015} and the development of local oscillators~\cite{HA¼bers2005}. 

In future, the Terahertz frequency range might offer a fascinating playground for quantum optics studies beyond the phenomena demonstrated at optical and microwave frequencies. In particular, the physics in the limit of ultra-strong and deep strong light matter coupling predicts light emission through non-trivial mechanisms~\cite{Ridolfo:2013bi,Ridolfo:2012dt}, such as from a non-excited system in its ground state which is non-adiabatically switched between uncoupled and ultra-strongly coupled regime. Notably, the technological maturity of nanoprocessing and epitaxial growth have allowed for above $\frac{\Omega_{r}}{\omega_{c}}=0.5$ coupling strengths to be demonstrated first at THz frequencies~\cite{Scalari:2012ia}~(0.58 at 500~GHz) and the frontier has been continuously pushed further to coupling strengths close to unity~\cite{Maissen:2014di}~(0.87 at 500~GHz). 

The quantum nature of photonic systems and the underlying emission mechanisms of light are reflected in the temporal correlation of their emission. The pioneering work of Hanbury Brown and Twiss~\cite{BROWN1956}, who raised big controversy with their measurement of coherence properties from star light, has nowadays essential technological implications in the characterization of light sources (single photon emitters, lasers, thermal radiation). Crucial requirements on the detectors including very low background noise, high sensitivity but also very good temporal resolution have seriously limited the experiments that could be performed so far at THz frequencies. 

In this paper, we exploit nonlinear parametric interaction between the THz field under investigation and femtosecond probing pulses for the measurement of field and intensity correlations at THz frequencies. We characterize our technique in comparison with state-of-the-art detector technology and highlight its advantages. We demonstrate good temporal resolution, high sensitivity and low background noise. With this technique, we measure the threshold statistics of a THz Quantum Cascade Laser.

\section{CORRELATION MEASUREMENTS at THZ FREQUENCIES}
Correlations are ubiquitous tools for the characterization of statistical properties of quantum fields. Two distinct quantities have been introduced for the description of light sources and their optical coherence: $g^{1}(\tau)$ representing the degree of first order coherence, and $g^{2}(\tau)$, the degree of second order coherence.

$$g^{1} (\tau)= \frac{ \langle \mathcal{E}^{-}_{THz}(t) \mathcal{E}^{+}_{THz}(t+\tau) \rangle_{t} }{\sqrt{\langle \mathcal{E}^{-}_{THz}(t)  \mathcal{E}^{+}_{THz}(t)  \rangle_{t} \langle \mathcal{E}^{-}_{THz}(t+\tau)  \mathcal{E}^{+}_{THz}(t+\tau) \rangle_{t}}}$$
 $$g^{2}(\tau) = \frac{ \langle \mathcal{E}^{-}_{THz}(t)\mathcal{E}^{-}_{THz}(t+\tau) \mathcal{E}^{+}_{THz}(t+\tau)\mathcal{E}^{+}_{THz}(t) \rangle_{t} }{\langle  \mathcal{E}^{-}_{THz}(t)  \mathcal{E}^{+}_{THz}(t) \rangle_{t} \langle \mathcal{E}^{-}_{THz}(t+\tau)  \mathcal{E}^{+}_{THz}(t+\tau) \rangle_{t}}.$$

The Hanbury Brown and Twiss experiment measures the degree of second order coherence. It is performed by splitting the radiation under investigation with a beam splitter, delaying one arm by an arbitrary time $\tau$ and measuring the coincidence of detecting photons at two distinct intensity detectors. The necessity for ultra-short detection times has been raised since its introduction, and the use of fast detection schemes~(100~fs) marked the point at which Hong, Ou and Mandel~\cite{PhysRevLett.59.2044} measured anti-bunching of photons. Alternative ways towards very short detection times have been proposed, among which two-photon absorption lead to the first measurement of photon bunching from thermal sources~(1~fs resolution)~\cite{Boitier:2009p1514}. 

Nowadays THz detector technology permits for detectors with relatively short detection times and good sensitivity\cite{Cai2014}\cite{Vicarelli:2012ch}. However, they are subject to current research and typically require cryogenic cooling for low background noise. Commercially available solutions are mostly based on inherently slow pyroelectric detectors (ms). An overview of selected THz detector technology is presented in table~\ref{tab:det} together with established NIR  technology for comparison.

\begin{table}[ht]
\caption{Response time of various state-of-the-art detectors compared to the detected frequency.} 
\label{tab:det}
\begin{center}       
\begin{tabular}{|l|l|} %% this creates two columns
%% |l|l| to left justify each column entry
%% |c|c| to center each column entry
%% use of \rule[]{}{} below opens up each row
\hline
\rule[-1ex]{0pt}{3.5ex}  Detector type & Response time //~in units of oscillation period ,$\frac{\tau_{resp}}{T_{det}}$ \\
\hline
\rule[-1ex]{0pt}{3.5ex}  Electro-optic sampling (2~THz) &  100~fs // $\approx$~0.2  \\
\hline
\rule[-1ex]{0pt}{3.5ex}  Graphene based hot electron bolometer~(1~THz) & 50~ps // $\approx$~50   \\
\hline
\rule[-1ex]{0pt}{3.5ex}  Pyroelectric detector~(1~THz) & 1~ms // $\approx$~1000000000 \\
\hline
\rule[-1ex]{0pt}{3.5ex}  Two-photon absorption (1550~nm)  & 2.5~fs // $\approx$~0.5 \\
\hline
\rule[-1ex]{0pt}{3.5ex}  Single photon APDs (1550~nm) & 20~ps// $\approx$~4000   \\
\hline
\end{tabular}
\end{center}
\end{table} 

\subsection{Electro-optic sampling for correlation measurements}
\begin{figure} [htb]
   \begin{center}
   \begin{tabular}{c} %% tabular useful for creating an array of images 
   \includegraphics[height=3.5cm]{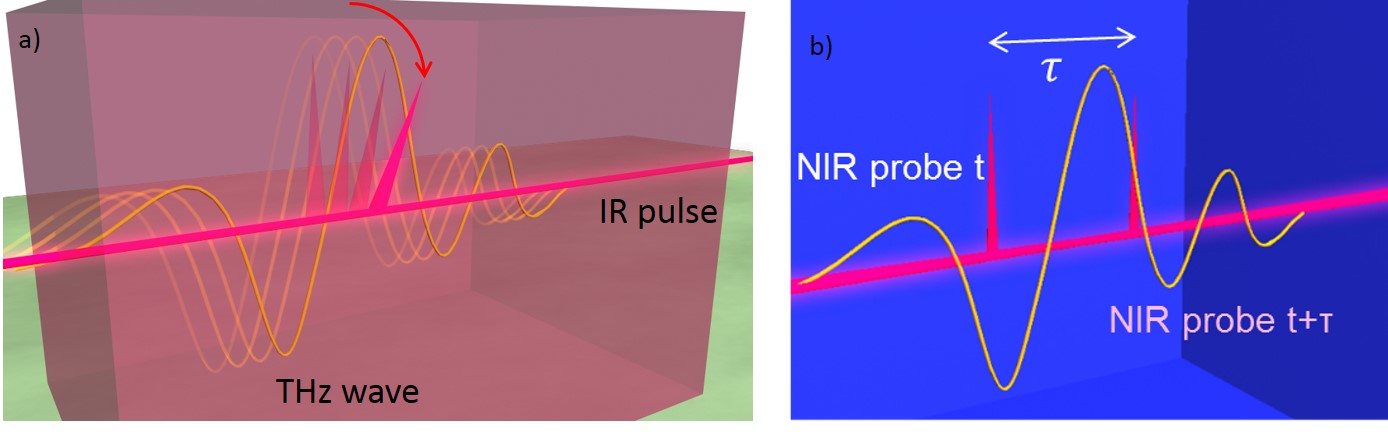}
   \end{tabular}
   \end{center}
   \caption[Principle of EOS and correlation] 
%>>>> use \label inside caption to get Fig. number with \ref{}
   { \label{fig:eoscorr} Principle of EOS and correlation measurements. a) In EOS, the polarization state of an ultra-short probe is altered locally by the presence of a THz wave, linearly to the magnitude to the electric field at the sampling point. b) The correlation measurement scheme is based on two mutually delayed probing beams which sample the THz wave at two distinct timepoints without the necessity of a beam splitter.}
   \end{figure}

  Among the coherent detection schemes, electro-optic sampling~(EOS) utilized for Terahertz Time Domain Spectroscopy, makes use of the parametric non-linear interaction between a THz wave and a femtosecond long probing beam inside a $\chi^{(2)}$ medium in a travelling wave configuration. Upon interaction, the polarization of the probing beam is altered linearly to the THz electric field magnitude~(figure~\ref{fig:eoscorr}a)). The interaction time is thus limited to the width of the probing pulse, which is typically on the order of 100~fs, rendering it very suitable for correlation measurements at ultrashort timescales. We exploit EOS to measure first and second order degree of coherence at THz frequencies. 
  
  A correlation measurement requires two distinct operations: the individual measurement of the same physical quantity at two distinct time-points and the averaging subsequent over all relevant times. While in standard techniques, the first requirement is met by use of a beam splitter and two distinct detectors, in our approach, we make use of a second, mutually delayed~(by a time $\tau$) probing pulse, combined with real-time measurement (at the laser repetition rate) of the electric field~(figure~\ref{fig:eoscorr}b)). The time averaging is performed digitally post-measurement. We used a frequency doubled Erbium Doped Fiber Amplifier~(EDFA) with a pulse width of 146~fs. A full description of the setup comprising all technical details is given elsewhere, in Ref.~\cite{2015arXiv151202198B}.

\begin{figure} [ht]
   \begin{center}
   \begin{tabular}{c} %% tabular useful for creating an array of images 
   \includegraphics[height=7cm]{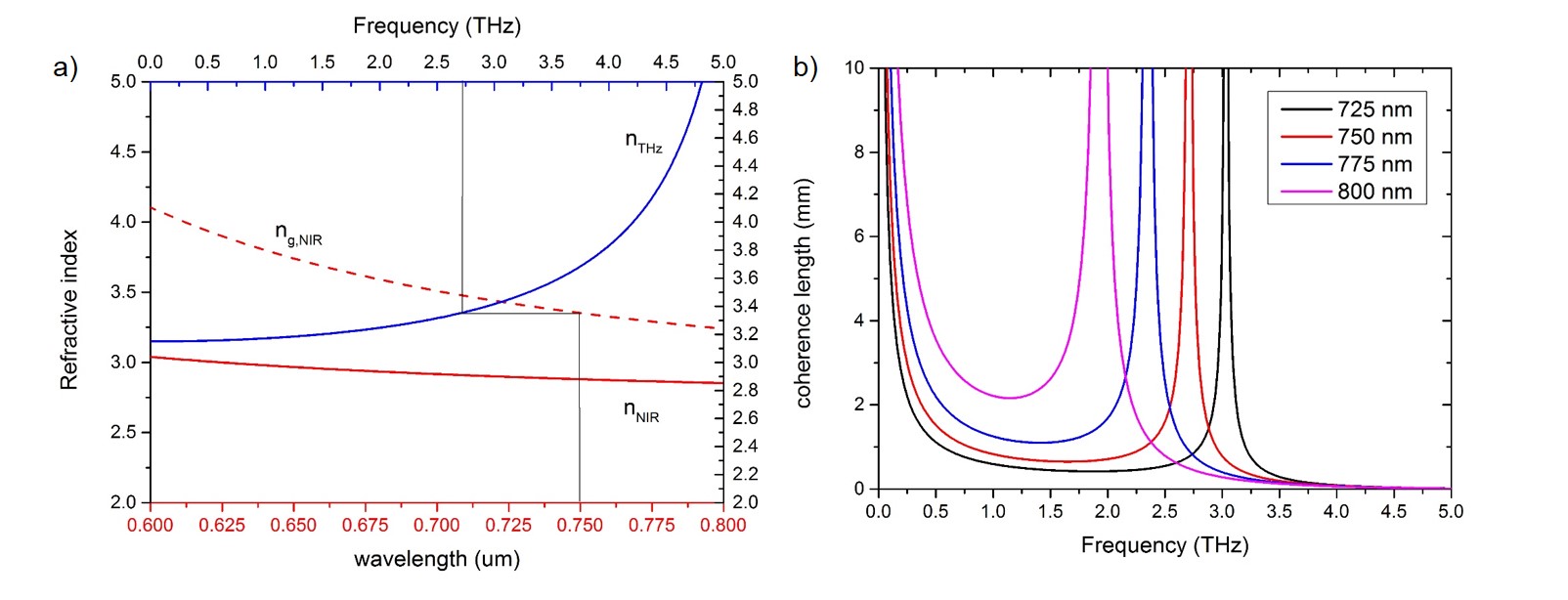}
   \end{tabular}
   \end{center}
   \caption[Principle of EOS and correlation] 
%>>>> use \label inside caption to get Fig. number with \ref{}
   { \label{fig:DetectionbandwidthZnTe} Coherence properties on 110-oriented ZnTe crystal in the detection range 0.1-3~THz. a) Refractive index of ZnTe at both interacting wavelengths. Schematic of coherence condition. b) Coherence length of ZnTe for different probe central wavelengths within the Ti:Sapph bandwidth.}
   \end{figure}

Electro-optic sampling presents various advantages over other detection schemes available for Hanbury Brown Twiss experiments at THz frequencies. Being a parametric coherent technique, the electro-optic effect is most efficient when the coherence condition is met: the phase index of the THz wave has to equal the group index of the probing pulse~\cite{Gallot:1999ty}\cite{Adam2007}. The interplay between probe wavelength, temperature and probe width allows for wide tunability and close-to-ideal coherence properties. The use of 110-oriented ZnTe as a detection crystal with $r_{41} = 4~pm/V$, together with tunable pulses from a Ti:Sapph laser~(center wavelength 725-850~nm), allows for coherence lengths on the order of several milimeters to be achieved anywhere in the range from 0.1-3~THz, as shown in figure~\ref{fig:DetectionbandwidthZnTe}.  For the computation of the coherence length we used the frequency dependent refractive index reported here~\cite{:/content/aip/journal/apl/69/16/10.1063/1.117511}. Unlike conventional detectors, the bandwidth of detection extends over several octaves.

The flexibility of this technique makes it applicable to other frequencies in the neighboring lower and higher frequency regions, with a fundamental Niquist limit imposed by the length of the femtosecond pulse. Nowadays technology permits for few fs pulses~(5~fs) to be reliably produced, which currently sets the limit to roughly 3~$\mu m$ radiation. In addition, this, or a similar technique might be an interesting option also for circuit QED experiments in replacing linear detectors which have an intrinsically narrow bandwidth~(few tens of MHz).

\section{PHOTON STATISTICS OF A THZ QUANTUM CASCADE LASER}

The photon statistics of a laser are expected to change at its threshold, from spontaneous emission with Superpoissonian statistics below threshold~($g^{2}(0)$~=~2), to coherent radiation above threshold with Poissonian statistics~($g^{2}(0)$~=~1). A Terahertz Quantum Cascade Laser~(QCL) is therefore a well suited device to be investigated in our setup. In addition, it has a typical bandwidth on the order of 20-30~$\%$, and the short temporal resolution of the correlation scheme we propose in this paper could be exploited to its full potential when performing this measurement below lasing threshold. 

We investigate the current dependent photon statistics of a THz QCL with an emission centered around 2.3~THz. The temporal resolution of our measurement of 146~fs is shorter than one cycle of oscillation of the THz emission, rendering it a sub-cycle measurement of $g^{2}(\tau)$. The laser is 1~mm long and cooled to 25~K inside a cryostat, and the measurements are performed at lab temperatures. For better understanding, we perform a theoretical analysis described in the following.
   
\subsection{Maxwell Bloch formalism for emission dynamics}

The theoretical model we use to describe the photon statistics and the intra-cavity photon field of the laser is based on a Maxwell-Bloch formalism combined with modal decomposition and is discussed in detail here~\cite{Villares:2015ho}.  The output describes the time-resolved evolution of all lasing modes at a given pumping, modeled as gain. It takes into account a Lorentzian gain profile, dispersion and four wave mixing. We use this model to predict the threshold statistics of the laser as well as a current-dependent analysis in the whole dynamic range. To do so, we add to the existing model the spontaneous emission, which is the limiting quantum noise and will have a major influence on the photon statistics below and during threshold. The complete description of the modal amplitudes concludes to 
\begin{eqnarray}
 \dot{A_n} = \left \{ \underbrace{G_n - 1}_{\hbox{\scriptsize{Net gain}}} +  i  \underbrace{\left( \frac{ \omega_n^2 - \omega_{nc}^2 }{ 2 \omega_n} \right )}_{\hbox{\scriptsize{Cavity dispersion}}} \right \} A_n  -   \\
  G_n \sum_{k,l}  
 \underbrace{ 
  C_{kl} B_{kl} A_{m} A_{l} \left \{ A_{k}^* \kappa_{n,k,l,m}    
  + A_{k-1}^* M_{+} \delta^* \kappa'_{n,k-1,l,m_{+}}  
+A_{k+1}^* M_{-} \delta \kappa'_{n,k+1,l,m_{-}} 
\right \}
}_{\hbox{\scriptsize{FWM term}}}  + \underbrace{A^{spont}_{n}}_{\hbox{\scriptsize{spontaneous emission term}}}
. 
\label{eq:modedynamicsall}
  \end{eqnarray}

Here $A_{n}$ is the complex amplitude of each mode in the laser cavity, $G_{n}$ the normalized gain, and $A_{n}^{spont}$ the Langevin force which describes the quantum fluctuation. It is modeled as a memory-less Markoffian random process with an amplitude corresponding to one single photon inside the laser cavity. The computation of $g^{2}(\tau)$ is trivial from the steady-state solution of the complex amplitudes.

\begin{figure} [htb]
   \begin{center}
   \begin{tabular}{c} %% tabular useful for creating an array of images 
   \includegraphics[height=9.2cm]{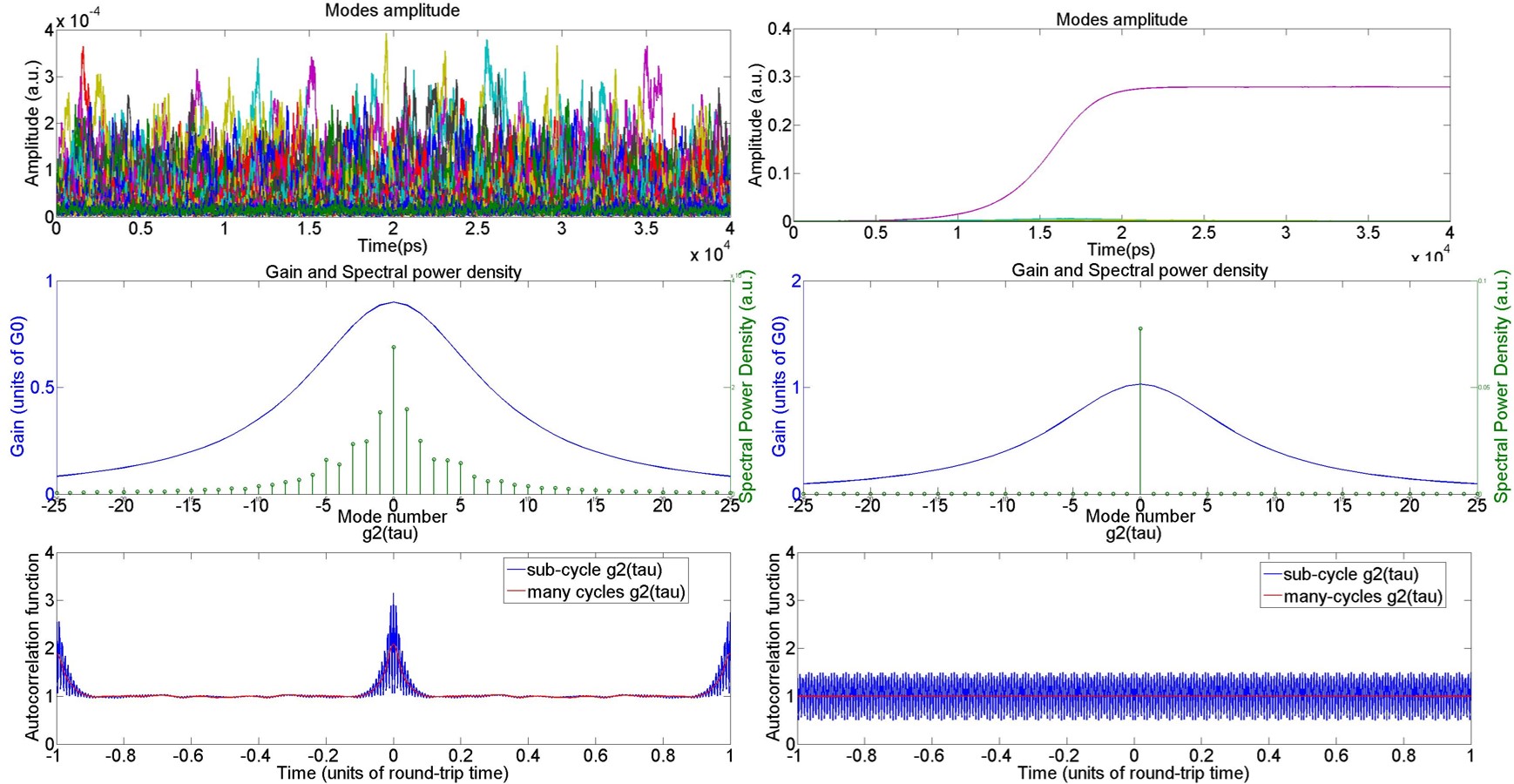}
   \end{tabular}
   \end{center}
   \caption[Principle of EOS and correlation] 
%>>>> use \label inside caption to get Fig. number with \ref{}
   { \label{fig:g2-comp} Time-resolved computations of the intracavity field and the expected sub-cycle and many-cycle outcome of a $g^{2}(\tau)$ measurement for the laser below threshold (a) and above threshold (b). The plots show the modal amplitudes, the gain and the spectral power density, and the simulated autocorrelation function (up to down). The specific photon statistics: bunched photons below threshold and a coherent state above threshold are retrieved.}
   \end{figure}

Figure~\ref{fig:g2-comp} shows characteristic temporal traces of the laser modes below~(a) and above threshold~(b), together with the modes supported by the laser cavity and the expected outcome of both a sub-cycle~(as in our case) and a many-cycle~(as in most cases) $g^{2}(\tau)$ measurement of the laser emission. 

The computation predicts various expected outcomes. Below threshold~(here 90$\%$ of threshold gain), the gain of the laser is smaller than the losses and only spontaneous radiation is present in the cavity. The intra-cavity power is very low~(small modal amplitudes) and decays at the cavity decay rate. The many-cycle value of $g^{2}(\tau = 0)$~=~2, as expected for spontaneous~(bunched) emission. Due to interference effects, the sub-cycle value of $g^{2}(\tau = 0)$~=~3.5. Above threshold~(here 103$\%$  of threshold gain), the laser is operating in single-mode since the gain is bigger than the losses only at its peak value. The modes inside the laser cavity experience gain and one gets amplified until reaching steady-state. The many-cycle value of $g^{2}(\tau = 0)$~=~1, as expected for lasing into one single coherent mode. The sub-cycle temporal shape of $g^{2}(\tau)$ represents a cosine wave with an amplitude of 0.5 and an average of 1. The periodicity of the cosine wave corresponds to twice the lasing frequency, since the instantaneous intensity of a coherent state oscillates at double the emission frequency. 

\subsection{Sub-cycle measurement of $g^{1}(\tau)$ and $g^{2}(\tau)$}
In the following we will present the results of the measurement of $g^{1}(\tau)$ and $g^{2}(\tau)$ with the electro-optic sampling method we present in this paper.

\begin{figure} [htb]
   \begin{center}
   \begin{tabular}{c} %% tabular useful for creating an array of images 
   \includegraphics[height=6cm]{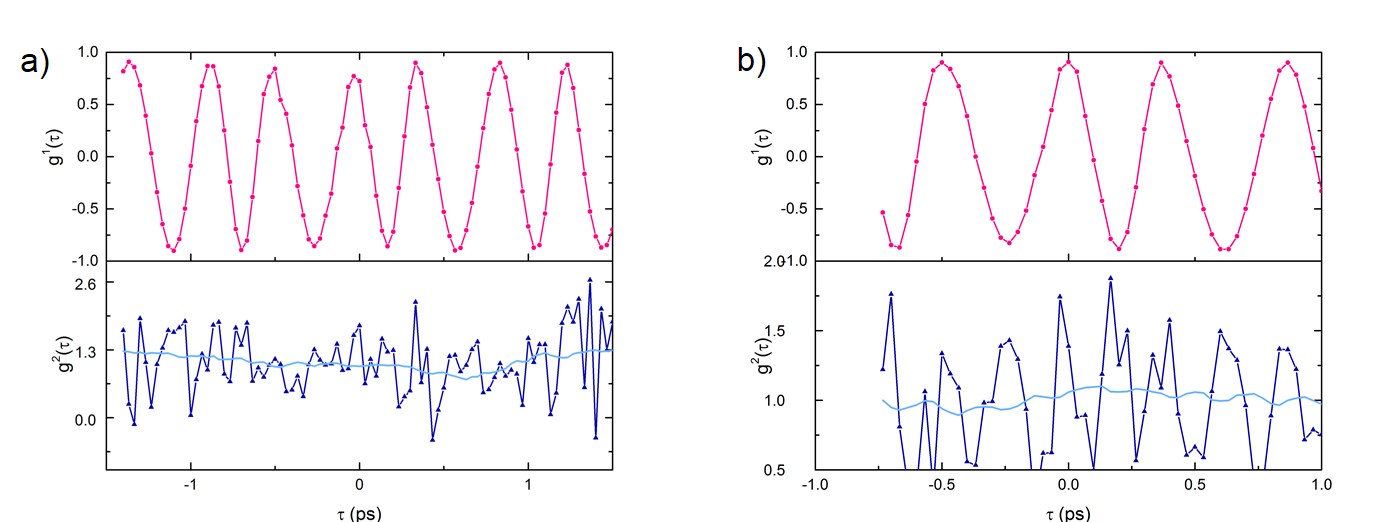}
   \end{tabular}
   \end{center}
   \caption[Principle of EOS and correlation] 
%>>>> use \label inside caption to get Fig. number with \ref{}
   { \label{fig:g2-meas}  Measurement results of $g^{1}(\tau)$ and $g^{2}(\tau)$ with fast electro-optic sampling in the threshold region~(a) and above threshold~(b), showing an increased value of $g^{2}(\tau = 0) = 1.18$ and $g^{2}(\tau = 0) = 1$, respectively. Double-frequency oscillations are retrieved for the sub-cycle measurement of $g^{2}(\tau)$.}
   \end{figure}

Figure~\ref{fig:g2-meas} shows characteristic results from the laser in the threshold region~(a) and above threshold~(b). We retrieve the expected lasing frequency around 2.3~THz from the measurement of $g^{1}(\tau)$, whose Fourier transform gives the spectrum of emission. In addition, we find an increased value of $g^{2}(\tau=0) = 1.18$ for the laser in the threshold region, as expected at this operating point. Above threshold, we retrieve a characteristic shape of $g^{2}(\tau)$ as expected: a cosine wave oscillating around 1 with an amplitude of 0.5 and double the emission frequency. 
  
Further reading concerning measurements and analysis of the sensitivity limit of the setup are extensively discussed here~\cite{2015arXiv151202198B}.

\section{OUTLOOK}
We have demonstrated the successful implementation of the first setup capable of measuring the first and second order degree of coherence at THz frequencies with unprecedented sub-cycle (146 fs) resolution. The properties of this type of setup, including very detection bandwidth, high tunability and high sensitivity make it a valuable solution in the Terahertz range and might be adapted for the neighboring mid-IR and microwave regions as well.

\section*{ACKNOWLEDGEMENTS}
The presented work is supported in part the ERC project MUSIC as well as by the NCCR Quantum Science and Technology.

% References
\bibliography{report} % bibliography data in report.bib
\bibliographystyle{spiebib} % makes bibtex use spiebib.bst

\end{document}